\documentclass[preprint,authoryear,12pt]{elsarticle}

\usepackage{graphicx}

\usepackage{amssymb}

\journal{Journal of Environmental Radioactivity}

\begin{document}

\begin{frontmatter}



\title{Aerial Measurement of Radioxenon Concentration off the West Coast of Vancouver Island following the Fukushima Reactor Accident}


\author{L.E.~Sinclair$^{1}$, H.C.J.~Seywerd$^{1}$, R.~Fortin$^{1}$, J.M.~Carson$^{1}$, P.R.B.~Saull$^{2}$, M.J.~Coyle$^{1}$, R.A.~Van~Brabant$^{1}$, J.L.~Buckle$^{1}$, S.M.~Desjardins$^{1}$, R.M.~Hall$^{1}$}

\address{$^{1}$Geological Survey of Canada, Natural Resources Canada, 601 Booth St, Ottawa, Ontario, K1A 0E8, Canada\\
$^{2}$Institute for National Measurement Standards, National Research Council, 1200~Montreal Rd, Ottawa, Ontario, K1A 0R6, Canada\\}

\begin{abstract}
In response to the Fukushima nuclear reactor accident, on March 20th, 2011, Natural Resources Canada conducted aerial radiation surveys over water just off of the west coast of Vancouver Island.  Dose-rate levels were found to be consistent with background radiation, however a clear signal due to $^{133}$Xe was observed.  Methods to extract $^{133}$Xe count rates from the measured spectra, and to determine the corresponding $^{133}$Xe volumetric concentration, were developed.  The measurements indicate that $^{133}$Xe concentrations on average lie in the range of 30 to 70~Bq/m$^3$.
\end{abstract}


\end{frontmatter}


\section{Introduction}
\label{sec:intro}
During March and April of 2011, in response to the Fukushima reactor accident, Natural Resources Canada conducted aerial and truck-mounted radiation surveys in the Victoria, Vancouver and Haida Gwaii areas and off the west coast of Vancouver Island.  Although radiation levels from Japan were predicted to be extremely low in these areas, dose rates were measured as a precautionary safety measure.  No increased levels of radiation were detected and the dose rates were found to be consistent with normal background radiation.

The first of the surveys, conducted on March 20th, 2011, was an airborne survey with the equipment installed inside a Pilatus PC-12 aircraft.   The survey was flown at approximately 250~m above mean sea level along the west coast of Vancouver Island.  This survey was flown several kilometres out to sea in order to increase sensitivity to man-made radiation by decreasing the count rate coming from terrestrial radiation.  A clear radioxenon signal was observed on this flight.  However, although the equipment was calibrated for total air kerma, no calibration for the measurement of absolute radioxenon concentration was available.  Furthermore, the emergency response procedure followed for these measurements entailed the use of an aircraft belonging to the Royal Canadian Mounted Police, for which a calibration had not been completed, complicating the correction of the data for aircraft attenuation.

In this paper, we present our method for extraction of $^{133}$Xe count rate from the airborne data collected on March 20th.  We then present an approach to convert those count rates into a volumetric plume concentration.  Finally, we present the measured $^{133}$Xe concentrations.

\section{Data Collection}
\label{sec:data}
\begin{figure}
   \begin{center}
   \begin{tabular}{c}
   \includegraphics[width=6cm]{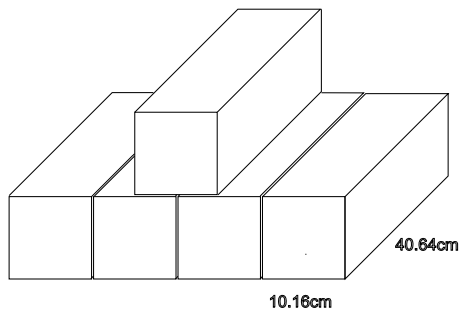}
   \end{tabular}
   \end{center}
   \caption
   { \label{fig:diagram} 
Geometry of the RSX-5 detector, showing the arrangement of the five large rectilinear NaI(Tl) crystals.}
   \end{figure}

Gamma spectra were collected using an RSI RSX-5 detector~\citep{RSX5}.  The detector consists of five 
10.16~cm x 10.16~cm x 40.64~cm (4~inch x 4~inch x 16~inch) NaI(Tl) crystals.  The five crystals are packaged such that four lie beside each other and the fifth lies on top, as indicated in Figure~\ref{fig:diagram}.
The crystals are optically isolated from each other and each is mated to a photomultiplier tube (PMT) for collection of the 
scintillation light.  Individual crystals are clad in Al and then wrapped in felt for cushioning.  The package assembly is 
constructed of foam and carbon fibre to provide structural support and insulation while allowing maximal penetration of 
low-energy gamma rays.

The PMT pulses are digitized and recorded as a 1024-channel energy spectrum.  Energies in the spectrum 
range from 0 to 3~MeV with one channel recording hits due to cosmic radiation which lies in the range of 3~MeV to 6~MeV.  An energy calibration with a $^{137}$Cs check source is performed when the instrument is powered on and thereafter the gain is automatically stabilized using the energy peaks due to naturally occuring isotopes.

A GPS receiver records position information and the GPS position and energy spectrum are recorded to a memory device (and optionally displayed on a computer in real-time) once per second.

On March 20th, 2011, the RSX-5 was installed in a Pilatus PC-12 fixed-wing aircraft.  The aircraft flew from Victoria, British Columbia, to the west coast of Vancouver Island and then surveyed moving north-west along the coastline a few kilometres out to sea.  The aircraft flew at a constant height of about 250~m above sea-level at an approximately constant speed of 410~km/hr to the north-west tip of the island.  On the return journey, the aircraft flew at a series of higher elevations, up to 2900~m.  Due to the complexity of correction of the higher altitude data for cosmic radiation effects, this paper will present only the data from the 250~m altitude outbound journey.  
The data presented here span a $\sim$~350~km distance and $\sim$~53~minute time interval.  To show the position-dependence of the data with sufficient statistical precision, the data have been accumulated into 20~s intervals.

On April 21st, 2011, over four weeks after the accident began, indications were that the radiation from Fukushima had reduced by several orders of magnitude~\citep{BGR,HC}.
A baseline survey was therefore conducted by flying along the coastline at 330~m altitude at an approximate speed of 400~km/hr.  To understand the cosmic contribution, additional data were collected at a series of different altitudes over the ocean off the north-west tip of the island.  

\section{Analysis}
\label{sec:analysis}
There are two major steps in the treatment of the data to obtain the activity of anthropogenic radionuclides: 1) the subtraction of expected natural backgrounds and 2) the conversion of count rates in s$^{-1}$ to volumetric concentration in Bq/m$^{3}$.

\subsection{Background subtraction}
\label{subsec:background}
\begin{figure}
   \includegraphics[width=14.3cm]{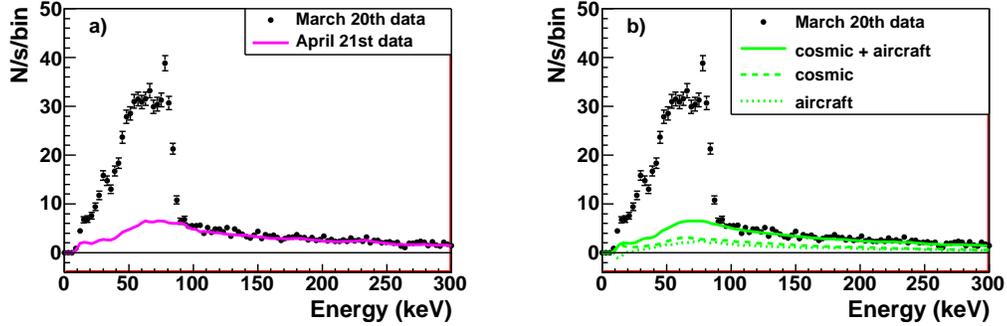}
   \caption
   { \label{fig:baseline} 
Measured energy spectrum for one 20~s interval of the data taken on March 20th, 2011 at 250~m above mean sea level
.  The error bars represent the one standard deviation statistical uncertainty.  In a) the curve shows the average measured spectrum from a baseline survey conducted along the Vancouver Island coastline on April 21st, 2011, where the baseline survey has been scaled to match the March 20th data over the energy range from 120~keV to 1500~keV.
In b) the solid curve shows an estimate for the cosmic and aircraft backgrounds derived from a geophysical survey conducted using the same kind of spectrometer but a different aircraft.  The dashed and dotted curves show the components of this background attributed to cosmic rays and to the aircraft, respectively.}
   \end{figure}

The measured spectrum from one 20~s interval, normalized to live-time, is shown in Fig.~\ref{fig:baseline}~a) by the black dots, where we have focussed on the energies below 300~keV.  A peak at energies around 80~keV and lower, is apparent in the data, sitting on top of a small natural continuum spectrum.

The spectrum from the baseline survey conducted April 21st, 2011, averaged over the entire flight length, is shown in 
Fig.~\ref{fig:baseline}~a) as the solid curve.  By examination of the relative heights of the $^{214}$Bi energy peak at 609~keV in the two datasets (not shown), it was determined that the radon concentration on April 21st was slightly lower than it had been on March 20th.  Therefore the baseline spectrum was scaled up by factors of $\sim$1.25 to match the data over the energy range of 120~keV to 1500~keV, for each 20~s interval of the March 20th data.

A clear excess above the baseline survey data is observed in the March 20th data, with a peak at an energy of 81~keV.  This is the photopeak energy of $^{133}$Xe, a known fission-product gas.  (A search was performed for other photopeaks due to fission products by accumulating the entire flight path into one spectrum, but no other significant contaminants were found.)

To obtain the count rate due to $^{133}$Xe, the scaled baseline survey spectrum was subtracted from the March 20th spectrum and the count rate in the difference spectrum was integrated over the energy range from 40~keV to 90~keV.  Of course, the $^{133}$Xe count rate presented herein is specific to our detector and method.  The determination of the actual plume concentration, which can be compared across measurement systems, is the subject of the following sub-section.

To understand the potential uncertainty introduced in the extracted $^{133}$Xe count rate by mis-measurement of the baseline spectrum, a number of alternate baseline spectra were examined.  It is common practice in geophysical surveying to establish the cosmic and aircraft backgrounds by taking a series of measurements at different altitudes over a body of water~\citep{IAEA323}.  For each spectral bin, a linear fit of the dependence of the count rate on counts in the cosmic channel is performed.  These slopes then allow for the interpolation of the cosmic component of the natural spectrum to any altitude, and the intercepts of the fit constitute the aircraft background.

Fig.~\ref{fig:baseline}~b) shows the same March 20th data sample, this time compared to cosmic and aircraft background spectra taken from a geophysical survey flown in 2007 over Eda Lake in the North West Territories.  This comparison survey used the same kind of RSX-5 survey system, but a different model of aircraft.  The $^{133}$Xe count rate extracted using this alternate baseline is in very good agreement with the signal extracted using the April 21st baseline flight.

Finally, a fit for cosmic and aircraft backgrounds was applied to data from flights taken at altitudes ranging from 2400~m to 6000~m off of the north-west tip of the island on April 21st.  The background spectrum obtained using these parameters is in excellent agreement with that obtained from the baseline survey along the coast.

In the presentation of count rates in Section~\ref{sec:results}, the variation in the signal obtained with the three different methods of background subtraction is included as a systematic uncertainty.

\subsection{Concentration conversion}
\label{subsec:concentration}
To understand the signal spectrum, and to obtain a conversion of count rate to concentration, we made a number of Monte Carlo simulations using the EGSnrc C++ code system~\citep{EGSnrc1,EGSnrc2}.

The RSX-5 simulation included the NaI(Tl) crystals in the Al cladding, the foam and felt between and around them, and the carbon fibre housing.

The aircraft was modelled as a uniform spherical shell of solid Al of nominal radius 2~m and varying thicknesses, with the RSX-5 situated at the centre of the spherical shell.

To obtain the effective thickness of the aircraft fuselage for the simulations, a series of attenuation measurements were conducted in the aircraft hangar on April 19th, 2011.  A $^{137}$Cs point source was placed at various locations against the outside of the aircraft body.  Counts due to this source were measured on the inside of the aircraft using a single 10.16~cm x 10.16~cm x 40.64~cm (4~inch x 4~inch x 16~inch) NaI(Tl) crystal and an RSI 705 spectrometer.  By comparison of the attenuated count rate with that expected in the absence of the aircraft, the attenuation due to the aircraft was measured.  The effective thickness of an equivalent uniform solid spherical shell of Al was then determined.  The aircraft showed considerable variation in effective thickness, being for example much thinner on the top than on the bottom, which is unsurprising.  The various thickness measurements were then combined in proportion to the relative area of the section of aircraft they represented and a rough estimate for average effective thickness of ($1.46 \pm 0.33$)~cm was obtained.

Geometric arguments based on the aircraft operating weight and overall dimension support this effective fuselage thickness, so the value of 1.5~cm was used in the nominal Monte Carlo simulations.  The estimate of aircraft effective fuselage thickness and the uncertainty arising from imperfect knowledge of it were further refined in the Monte Carlo study described below.

The source used in the simulations was spherical, centred on the RSX-5, and filled with some concentration of isotropically emitting point sources of gamma radiation.  The emitters were uniformly distributed throughout the spherical plume, both inside and outside of the simulated aircraft.  The nominal plume radius was 300~m.  The source spectrum representing $^{133}$Xe contained two lines, an 81~keV line with 38\% branching ratio, and a 31~keV line to represent all of the major X-ray emissions, with 50\% branching ratio.

The spherical plume of $^{133}$Xe was situated inside a cubic volume of air with sides of length four times the radius of the plume.  Interactions in the air inside and outside of the aircraft were simulated.   The air was treated as dry air at atmospheric pressure, and pressure differences between the inside and outside of the aircraft were neglected.

In these EGSnrc simulations, the XCOM~\citep{XCOM} photon cross sections were used and Rayleigh scattering and atomic relaxations were turned on.  Potential contributions to the spectrum from $^{133}$Xe beta decay and internal conversion processes were investigated but found not to warrant inclusion.
Electrons and photons were transported down to a kinetic energy of 10 keV.  (Going to lower energy, 1~keV, did not have a noticeable effect on the results, but increased the simulation time considerably.)  A total of 10$^{11}$ emissions were generated in the nominal configuration, leading to $\sim$15,000 counts interacting in the detector.  The statistical uncertainty on the generated spectra is smaller than that of the measured spectrum over a 20~s interval.

The simulated energy deposits in the NaI(Tl) crystals were smeared according to a resolution function for the RSX-5 which was tuned at low energies by measuring the photopeak widths for $^{241}$Am (59~keV), $^{109}$Cd (88~keV) and $^{57}$Co (122~keV).

\begin{figure}
  \includegraphics[width=7cm]{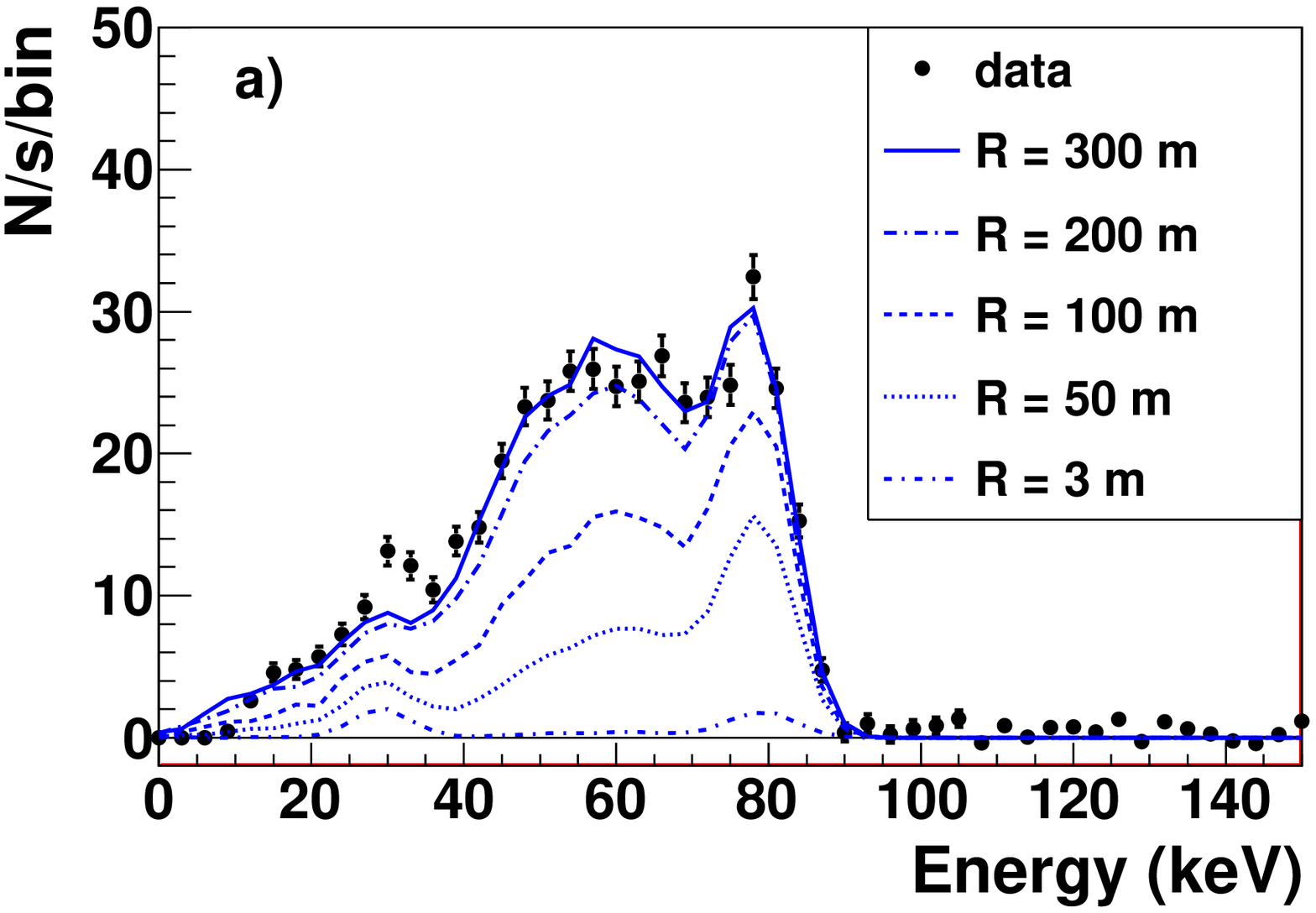}\includegraphics[width=7cm]{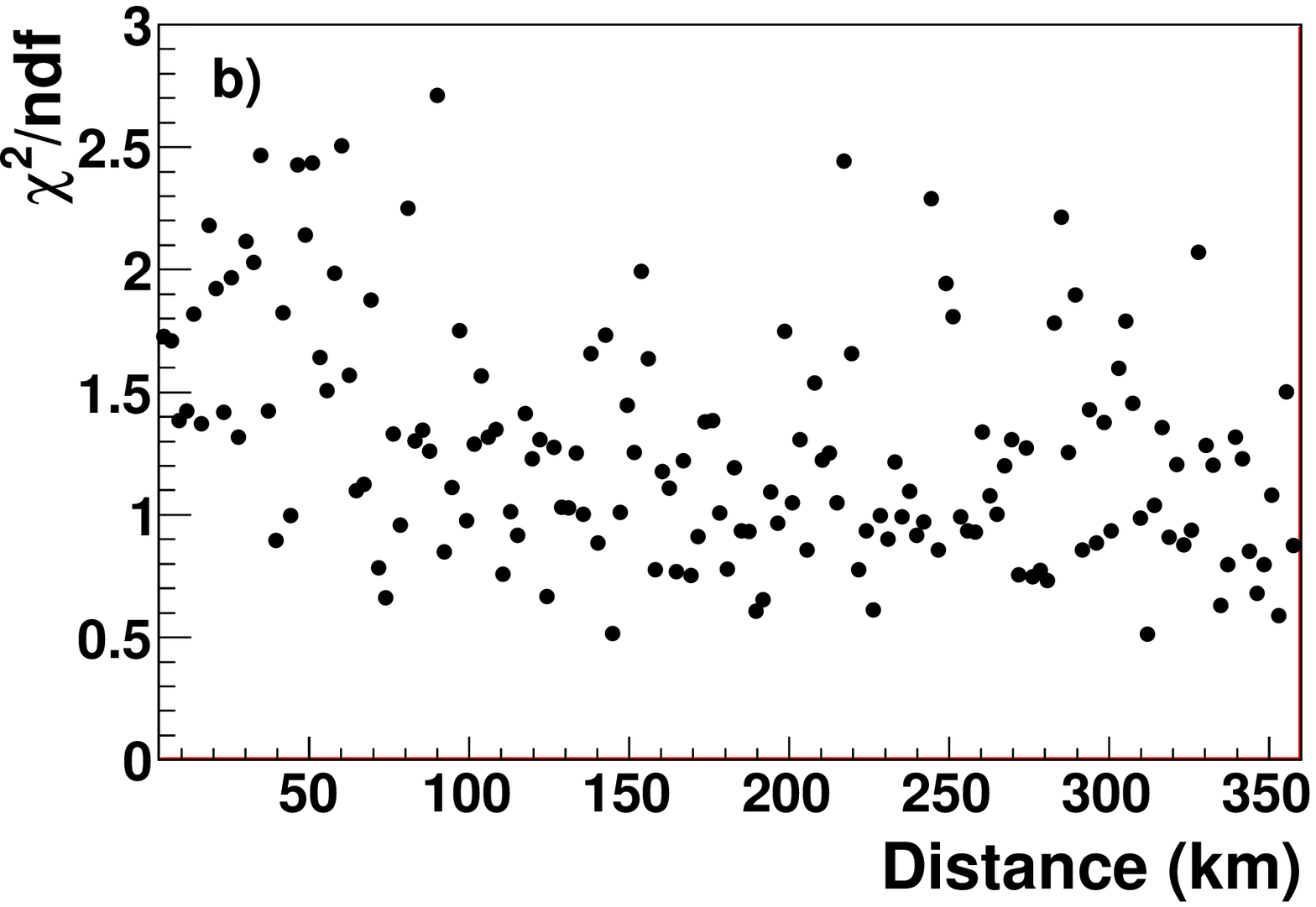}
  \includegraphics[width=7cm]{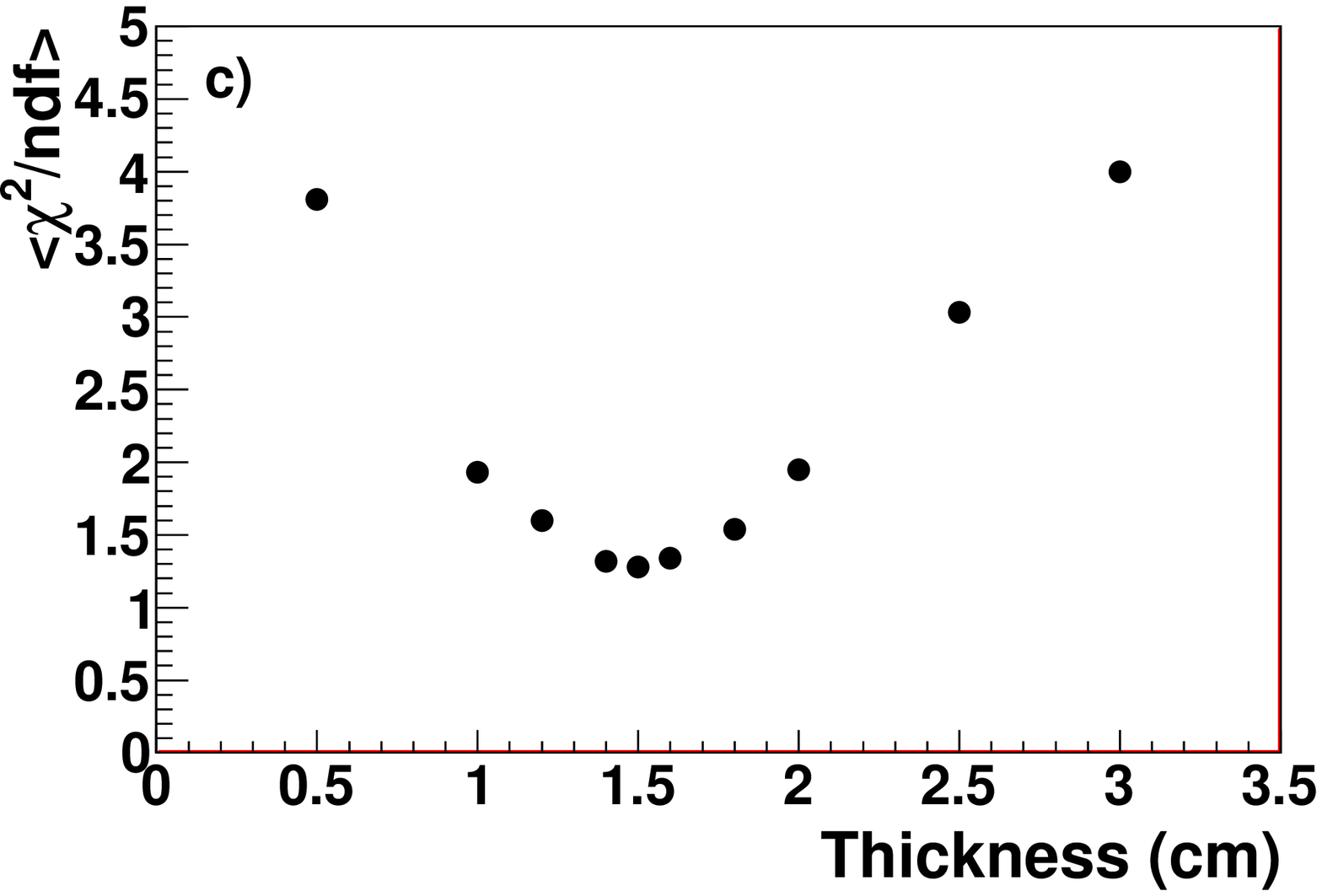}
  \caption 
  { \label{fig:sensitivity} 
In a) the black dots show the energy spectrum for the extracted signal for one 20~s interval of data.  The error bars represent the statistical uncertainty on the data.  The solid curve shows the Monte Carlo simulation for a plume of radius 300~m.  Additional curves show the Monte Carlo predictions for plumes of radii 200~m, 100~m, 50~m and 3~m, where all Monte 
Carlo predictions have been normalized to a plume concentration of 66~Bq/m$^3$.
In b) the $\chi^2$/ndf from the comparison of Monte Carlo to data is shown for each 20~s dataset as a function of distance along the coastline.
In c) the average $\chi^2$/ndf over all of the 20~s intervals is shown for different effective fuselage thicknesses used in the simulation.}
\end{figure}

In Fig.~\ref{fig:sensitivity}~a) the extracted $^{133}$Xe spectrum for one 20~s interval of the March 20th data is shown by the black dots.  The solid curve shows the prediction for the spectrum due to a $^{133}$Xe plume of 300~m radius obtained by simulation.  We find that the data are well described by the 300~m plume prediction, in the energy range of 40~keV to 90~keV.  

One may note that there is a discrepancy between data and simulation in the vicinity of the 31~keV X-ray emissions.  While the 81~keV gammas originate predominantly outside of the aircraft, attenuation by the aircraft body assures that any detected 31~keV X-rays were likely to have been emitted from within the aircraft.  The simulation of the X-rays is thus highly sensitive to details of the shape of the aircraft in the simulation.  This is also the energy where there are unsimulated lower-level discriminator effects in the data and where the details of the simulation of the detector housing become relevant.  
The excess at 31 kev could also hint at the presence of $^{\mbox{\scriptsize 131m}}$Xe or $^{\mbox{\scriptsize 133m}}$Xe which yield 
X-rays in the 30-35 keV region in more than half of their decays. 
(The 164~keV gamma of $^{\mbox{\scriptsize 131m}}$Xe is emitted in only 2\% of its decays and the 232~keV gamma of 
$^{\mbox{\scriptsize 133m}}$Xe is emitted in only 8\% of its decays and the presence of these peaks in the data can not be ruled out.)

In the following our analysis is based entirely on the energy spectrum above 40~keV.

Fig.~\ref{fig:sensitivity}~a) also shows Monte Carlo predictions for spectra due to plumes of various smaller radii, down to a plume of 3~m radius, where all of the Monte Carlo curves have been normalized to the same plume concentration.  
The smaller plume radii illustrate how the scattering of the 81~keV gamma-rays in air leads to the population of the spectrum above the 31~keV X-ray region and below the 81~keV photopeak.  We find that plumes with radii less than 200~m lead to a spectrum which is inconsistent in shape with the measured $^{133}$Xe spectrum.  
We also determined (not shown) that generation of plumes with radii greater than 300~m did not improve our understanding of the data.  Due to attenuation in air, plumes with radii greater than 300~m are not significantly different in shape or magnitude from the 300~m radius plume.  Therefore, the data from this 20~s interval are consistent with emission from an infinite and uniform $^{133}$Xe plume.

The Monte Carlo simulations provide a $^{133}$Xe count rate to volumetric concentration conversion factor of 
(0.177 $\pm$ 0.001)~$\frac{\mbox{Bq/m$^3$}}{\mbox{s$^{-1}$}}$ (statistical uncertainty only).  Multiplying the measured count rate over the 40~keV to 90~keV interval by this conversion factor, the concentration of $^{133}$Xe for this 20~s interval was determined to be ($66.0 \pm 1.0$)~Bq/m$^3$ (statistical uncertainty only).  The concentrations for each of the other 20~s intervals were likewise determined by application of the count rate to concentration conversion factor.

To determine the goodness of each of the comparisons of Monte Carlo and data, the $\chi^2$ per degree of freedom ($\chi^2/\mbox{ndf}$) was calculated for each sample.  The $\chi^2/\mbox{ndf}$ values are shown in Fig.~\ref{fig:sensitivity}~b) as a function of distance along the coastline.  Overall, the simulation provides a good representation of the data, and no location stands out as having differently shaped spectra.

To quantify the uncertainty on the effective fuselage thickness, Monte Carlo samples were generated with thicknesses ranging from 0.5~cm to 3~cm.  The average 
$\chi^2$/ndf over all of the 20~s intervals, $<\chi^2/\mbox{ndf}>$, is shown in Fig.~\ref{fig:sensitivity}~c) versus the fuselage thickness used in the simulation.  By examination of Fig.~\ref{fig:sensitivity}~c) we have determined that an aircraft fuselage of 1.5~cm provides a good representation of the data, supporting the choice of 1.5~cm as the nominal effective fuselage thickness.  The effect on the calculated concentrations of variation of the fuselage thickness between 1.2~cm and 1.8~cm is included in the systematic uncertainty on the measured concentrations.

\section{Results}
\label{sec:results}
\begin{figure}[p]
  \rotatebox{90}{
    \begin{minipage}{\textheight}%
      \includegraphics[height=19.cm,angle=270]{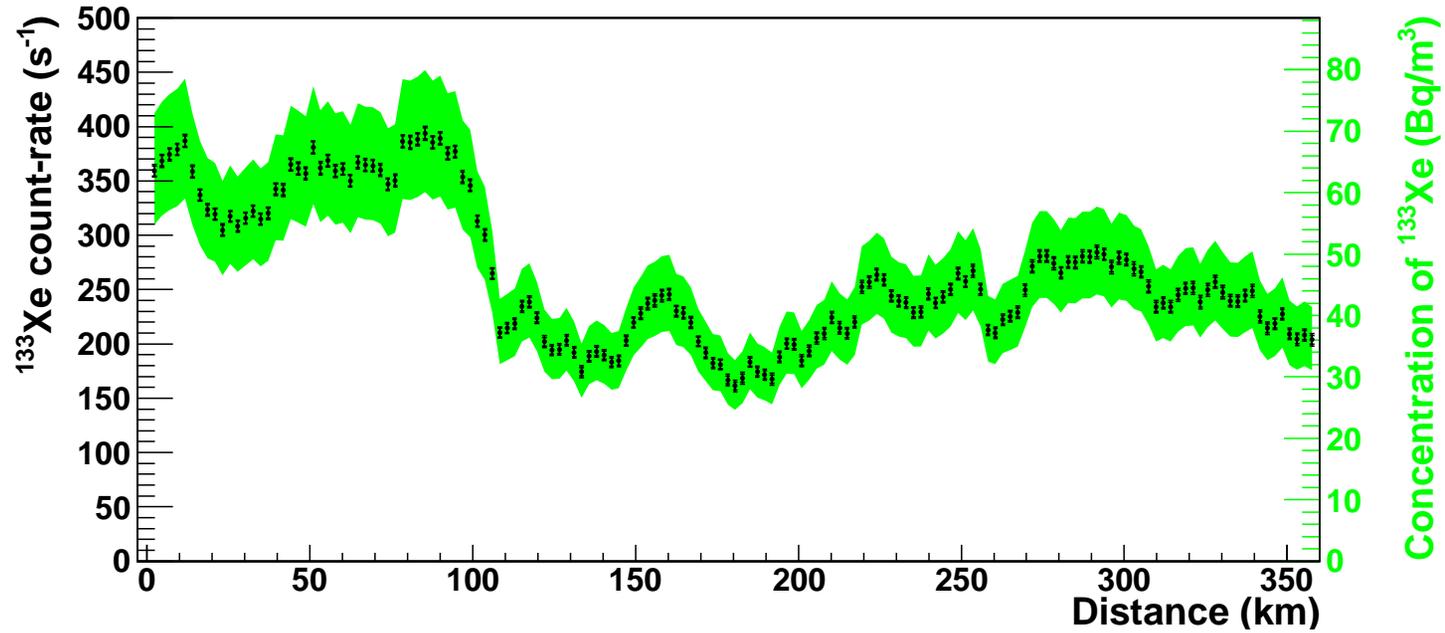}
      \caption{
        \label{fig:conc_vs_dist} 
        The measured $^{133}$Xe count rate in the interval of 40~keV to 90~keV as a function of distance from the survey starting point is shown by the black dots, read from the left-hand axis.  The inner error bars show the one standard deviation statistical uncertainty and the outer error bars show the systematic uncertainty associated with the background subtraction added in quadrature (not always distinguishable from the inner error bars).  Using the axis on the right-hand side, the measurements may be read as concentrations due to an infinite uniform plume of $^{133}$Xe.  The shaded band shows the correlated systematic uncertainty on the concentrations due to uncertainty in the effective aircraft fuselage thickness.}
    \end{minipage}%
  }
\end{figure}

In Fig.~\ref{fig:conc_vs_dist} the count rate due to $^{133}$Xe for each 20~s interval is shown as a function of distance along the coastline from an arbitrary starting point.  The time of the beginning of the first measurement at zero distance is March 20th, 2011, 11:02:58~am~PDT and each subsequent measurement is 20~s later.  Inner error bars show the one standard deviation statistical uncertainty and outer error bars show the systematic uncertainty due to the baseline subtraction added in quadrature.  (The systematic uncertainty is in many locations very small such that the outer error bars are not distinguishable from the inner error bars.)

We find that the signal due to $^{133}$Xe is significantly above zero at all locations.  
To our knowledge, this is a first publication of radioxenon concentrations measured by flying through the plume.  A notable flight through the Chernobyl plume was made by the Geological Survey of Finland, however that analysis was hampered by contamination of the aircraft due to the presence of radioactive particulates close to the plume origin~\citep{plume_cowboys}.
We also find that the count rate can vary strongly with position, by almost a factor of two over a 15~km distance, a flight taking just over two and a half minutes.  
Note that our simulations indicate that over 98\% of the $^{133}$Xe signal originated from sources outside of the aircraft.  Therefore each measurement reflects the $^{133}$Xe concentration at the local geographic coordinates.
The strong variability of count rate along the transection seems surprising given the distance the plume has travelled from Japan to Canada and it is hoped that this information will eventually prove useful in refining models for atmospheric transport.

Using the right-hand axis in Fig.~\ref{fig:conc_vs_dist}, the measurements may be read as plume concentrations.  The shaded band shows the correlated uncertainty on the concentration measurements coming from the uncertainty on the effective aircraft fuselage thickness.
We find that concentrations vary from about 30~Bq/m$^3$ to 70~Bq/m$^3$ along the coastline.  These measurements are in broad agreement with ground-based $^{133}$Xe measurements made at a different location following the Fukushima accident which have been published by another group~\citep{Bowyer2011}.  Given the very different technology and methodology used in the two analyses, the similarity of the results is a strong confirmation of the earlier measurement.

Note that we have made the assumption that the plume is infinite and uniform in extracting plume concentrations, while we clearly observe a variation in the plume concentration with distance.  The distance over which the  measurements in Fig.~\ref{fig:conc_vs_dist} were integrated was approximately 2.3~km.  In fact, due to the attenuation of air, the plume need only be uniform over distance scales of around 200~m in order for our concentration calculation to be valid.  Effectively, we have made the assumption that while the plume concentration can vary significantly over a distance of 2.5~km, the variation is gradual over distances of $\sim$~200~m.

\begin{figure}
   \includegraphics[height=12cm]{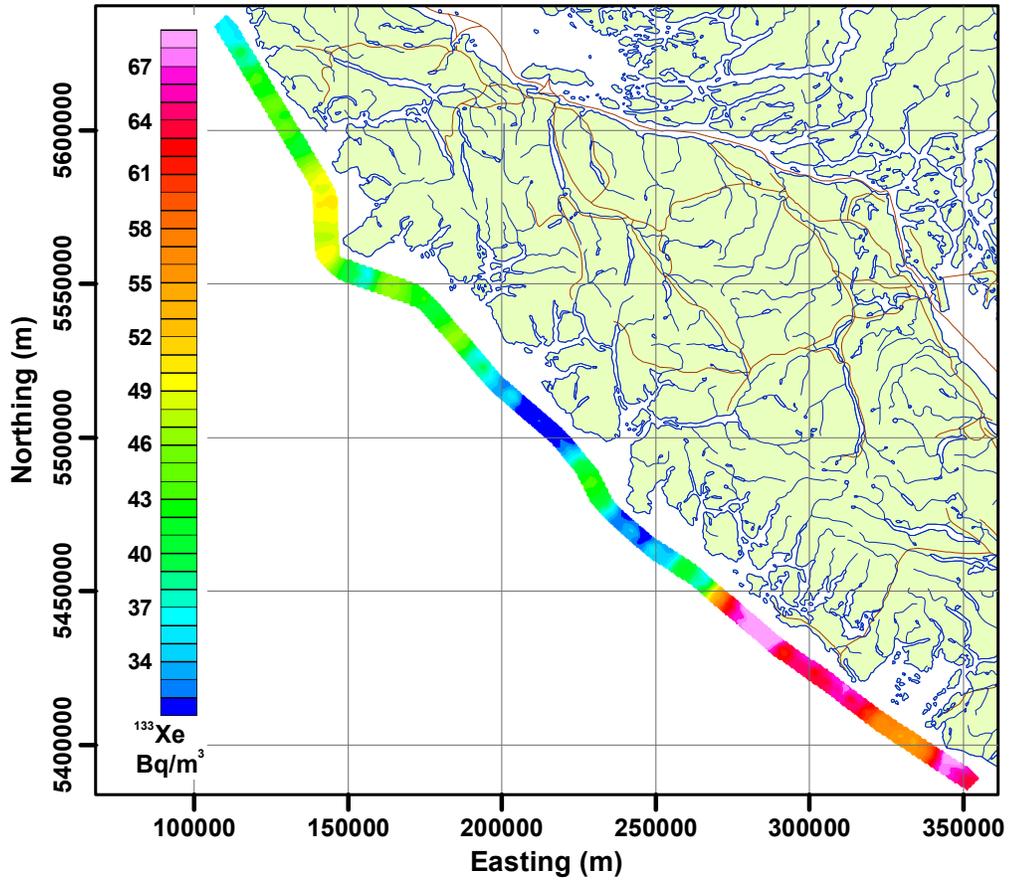}
   \caption[fig:Xe133_map] 
   { \label{fig:Xe133_map} 
$^{133}$Xe concentration is shown using a colour scale, versus the geographic coordinates Easting and Northing in Universal Transverse Mercator zone 10.}
   \end{figure}

A map showing the georeferenced concentration data overlaid on the local geography is presented in Fig.~\ref{fig:Xe133_map}.
It is anticipated that this information will be useful in validating calculations of the Fukushima reactor radioxenon inventories.

\section{Conclusions}
On March 20th, 2011, Natural Resources Canada flew radiometric surveys off the west coast of Vancouver Island, to investigate a suspected radioactive plume from the damaged Fukushima reactors.  We found a signal due to $^{133}$Xe and no significant signals due to other contaminants.  The $^{133}$Xe count rate was observed to vary strongly with position, by almost a factor of two over a distance of $\sim$~15~km.  Monte Carlo simulations were performed in order to understand the signal and convert it to a measurement of concentration.  The signal was found to be consistent with an infinite plume, uniformly populated with isotropic $^{133}$Xe point sources.  The concentration of the plume was determined to lie in the range of 30~Bq/m$^3$ to 70~Bq/m$^3$.  A map showing the variation of the $^{133}$Xe plume concentration with geographic position is presented.  This information has been used to support radioxenon measurements using alternate methods, and may be used to validate models for the progression of the Fukushima reactor accident.

\section*{Acknowledgements}

We are grateful to Dr.~J.~Hovgaard from Radiation Solutions Inc.\ and to K.~Ungar and E.~Korpach from Health Canada for scientific advice.  We gratefully acknowledge the Royal Canadian Mounted Police for their partnership and assistance with the mission.  Some of the funding for equipment used in this measurement was obtained from Canada's Chemical Biological Radiological and Nuclear Research and Technology Initiative.  This report is ESS Contribution Number 20110080.

\bibliographystyle{elsarticle-harv}
\bibliography{NRCan_Xe133}







\end{document}